\documentclass{ws-procs961x669}            

\usepackage{graphics,graphicx}
\usepackage{dcolumn}
\usepackage{bm}
\usepackage{epsfig}
\usepackage{mathbbol}
\usepackage{mathtools}
\usepackage{epstopdf}
\usepackage{simplewick}
\usepackage[utf8]{inputenc} 	
\usepackage{slashed}
\usepackage{stackrel}
\usepackage[normalem]{ulem}

\begin{document}
\title{Radiative processes and jet modification at the EIC}

\author{Ivan Vitev}

\address{Los Alamos National Laboratory, Theoretical Division, Mail Stop B283\\
Los Alamos, NM 87545, USA\\
$^*$E-mail: ivitev@lanl.gov\\
}

\begin{abstract}
A U.S.-based Electron-Ion Collider will provide the ultimate capability to determine both the structure and properties of nucleons and nuclei, as well as 
how matter and energy can be transported through a strongly interacting quantum mechanical environment.  The production 
and propagation of long-lived heavy subatomic particles is a unique and critical part of this planned decade-long research program.
In these proceedings we report the derivation of all branching processes  in nuclei that  lead to a modification of semi-inclusive hadron production, jet cross sections, and  jet substructure  when compared to the vacuum. This work allows for their evaluation to any desired order in opacity. As an example, we show an application to the modification of light hadron and open heavy flavor fragmentation functions at the EIC. We discuss how this observable can shed light on the physics of hadronization and parton energy loss in large nuclei.
\end{abstract}


\bodymatter

\section{Introduction}
The EIC is the top priority for new construction for DOE Office of Science, Office of Nuclear Physics and is expected to come online around 2028.
There is a short window of opportunity to sharpen its physics program and ensure that the experimental capabilities to carry out the required measurements  are available. To understand how partons propagate in cold nuclear matter and how hadrons form in the QCD environment is  an important pillar of the EIC science portfolio. Existing measurements from the HERMES Collaboration~\cite{Airapetian:2007vu}  have not been able to differentiate between competing models of light hadron attenuation~\cite{Kopeliovich:2003py,Arleo:2003jz} in semi-inclusive deep inelastic scattering (SIDIS) on nuclear targets. As we will show below, $D$-mesons and $B$-mesons production in such reactions is a sensitive probes of the nuclear matter transport properties and the suppression patterns these heavy particles  show are distinctly different for different theoretical models. This, in turn,
requires better understanding of medium-induced branching processes for light partons and heavy quarks~\cite{Kang:2016ofv} and the process of
hadronization.  
 
 \section{Radiative processes in DIS on large nuclei}
\begin{figure}[t]
\begin{center}
\includegraphics[width= 3.5in]{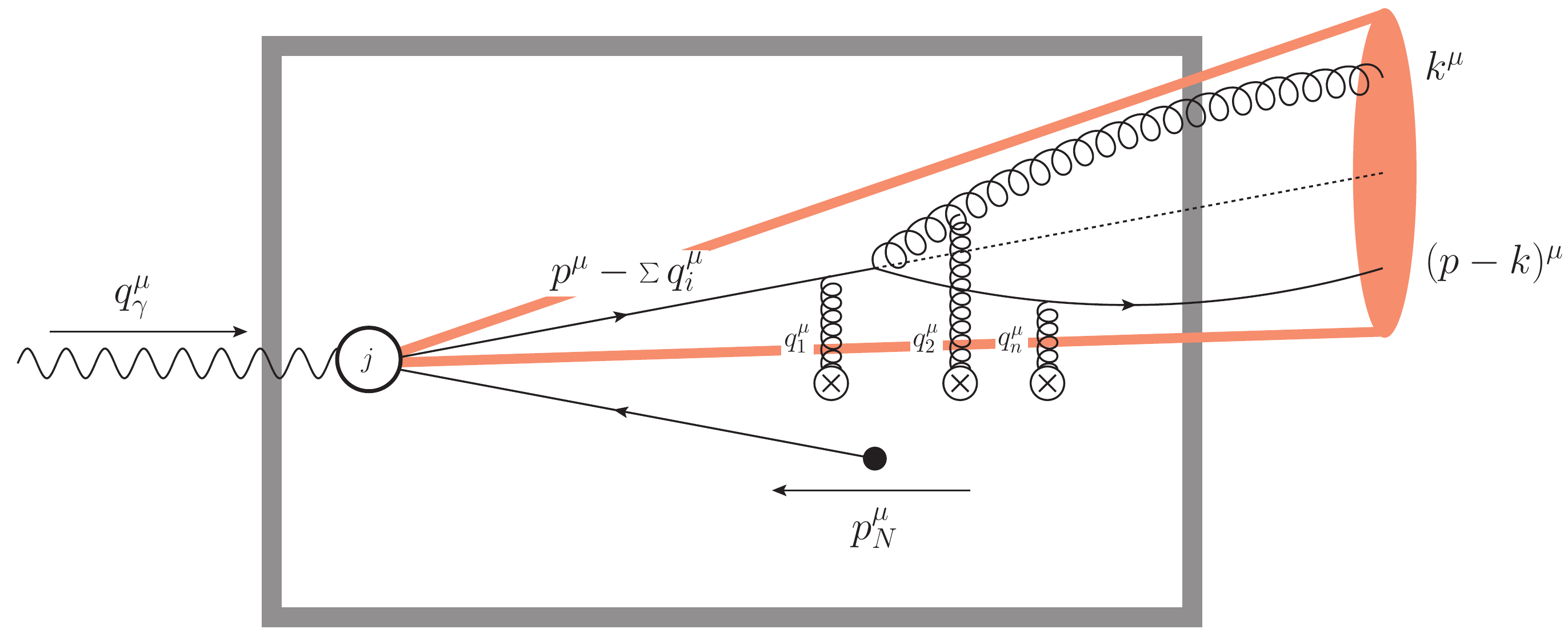} 
\caption{Illustration of the indicated jet kinematics for SIDIS in the Breit frame.  The dark box represents the medium (nucleus) and the red cone represents the jet.}
\label{f:Jet_Kinematics}
\end{center}
\end{figure}

Recently, we developed a formalism~\cite{Sievert:2018imd} that allows us to compute the gluon spectrum of a quark jet  to an arbitrary order in opacity, the average number of scatterings in the nuclear medium.  This calculation goes beyond the simplifying limit in which the gluon radiation is soft and  significantly extends previous work, which computes the full gluon spectrum only to first order in opacity, see Fig.~\ref{f:Jet_Kinematics}.  As all branching processes in matter have the same topology, the calculation can be generalized to obtain the full set of in-medium splittings~\cite{Sievert:2019cwq}.
The parton flavor and mass dependence are elegantly captured by the lightcone wavefunction formalism.

We can express the  medium-induced radiative processes as a sum of Initial/Initial,  Initial/Final,  Final/Initial and Final/Final
contributions, where the nomenclature refers to an interaction with a scattering center in matter before or after the splitting in the
amplitude and the conjugate amplitude, respectively. By evaluating Feynman diagrams corresponding to an interaction between the propagating parton system and the medium we derive both the initial conditions and kernels that relate $N$ and $N-1$  orders in opacity. 
The recursion relations between these four functions can be cast in the form of a matrix equation, with a particularly simple 
triangular form due to their causal structure:
\begin{align} \label{e:reactmtx}
\begin{bmatrix*}[l]
f_{F / F}^{(N)}  \\
f_{I / F}^{(N)}  \\
f_{F / I}^{(N)}  \\
f_{I / I}^{(N)} 
\end{bmatrix*}
= \int\limits_{x_0^+}^{\min[ x^+ , y^+ , R^+ ]} \frac{dz^+}{\lambda^+} 
\int\frac{d^2 q}{\sigma_{el}} \frac{d\sigma^{el}}{d^2 q}
\begingroup
\renewcommand*{\arraystretch}{1.4}
\begin{bmatrix}
\mathcal{K}_1^{} & \mathcal{K}_2 & \mathcal{K}_3 & \mathcal{K}_4 \\
0 & \mathcal{K}_5 & 0& \mathcal{K}_6 \\
0 & 0 & \mathcal{K}_7 & \mathcal{K}_8 \\
0 & 0 & 0 & \mathcal{K}_9
\end{bmatrix}
\endgroup
\begin{bmatrix*}[l]
f_{F / F}^{(N-1)}  \\
f_{I / F}^{(N-1)}  \\
f_{F / I}^{(N-1)}  \\
f_{I / I}^{(N-1)} 
\end{bmatrix*} ,
\end{align}
where the matrix of integral kernels $\mathcal{K}_{1-9}$ is an explicit representation of the reaction operator, $\lambda$ is the 
parton scattering length, and  ${d\sigma^{el}}/{d^2 q}$ describes the distribution of transverse momentum transfers.

We have written a Mathematica code to solve the recursion relation Eq.~(\ref{e:reactmtx}) and obtained explicit results to second 
order in opacity~\cite{Sievert:2018imd}. The fully differential spectrum of in-medium branchings has been evaluated 
numerically~\cite{Sievert:2019cwq}.

\section{Modification of light and heavy flavor mesons in SIDIS on nuclei}
\begin{figure}[h]
\begin{center}
\includegraphics[width=3.in]{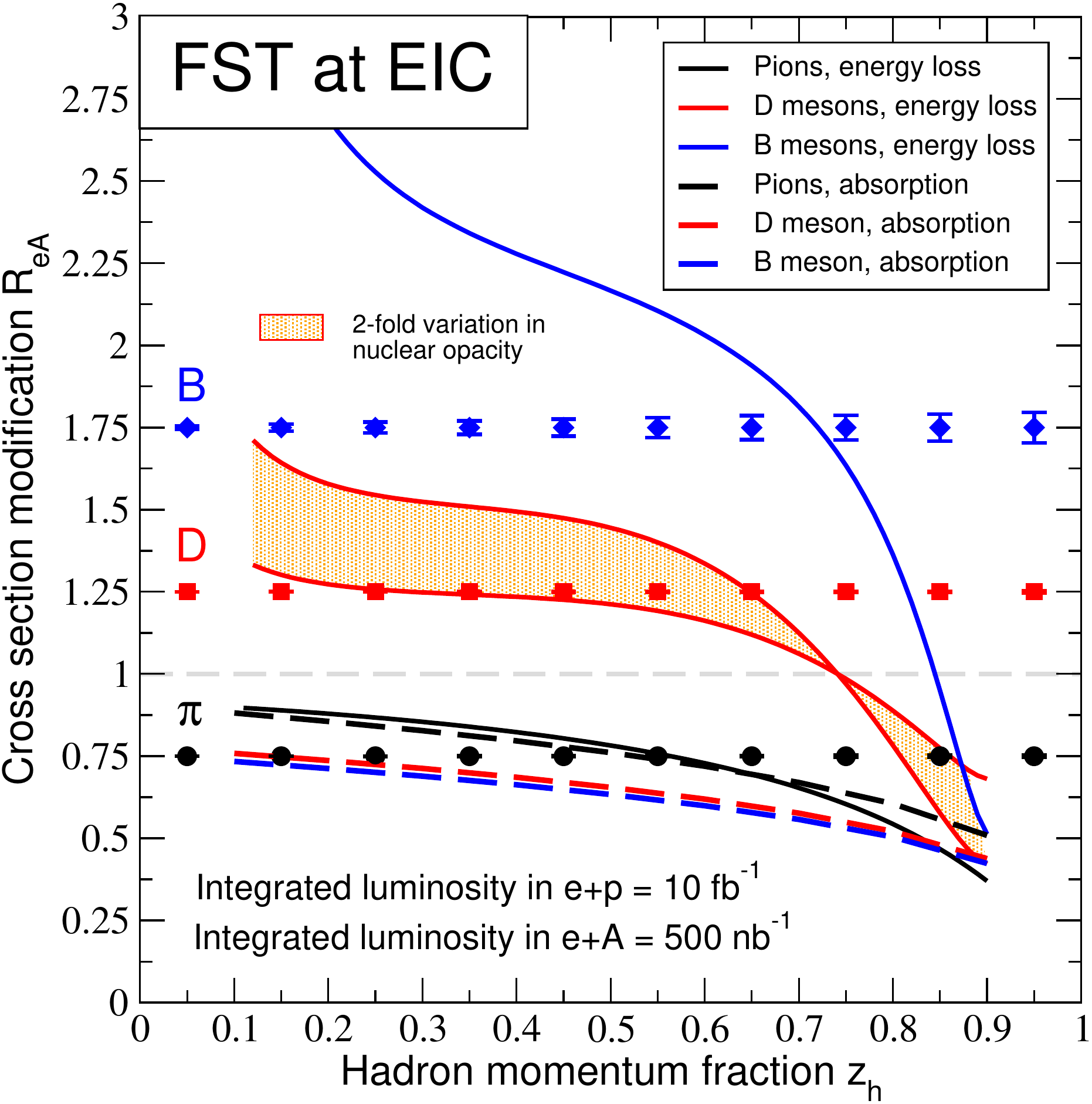}
\end{center}
\caption{Heavy quarks ($B$- and $D$-mesons) provide unparalleled discrimination power for the nature of relativistic transport through nuclear matter. Expected statistical precision for $\pi^\pm$ (black), $D^\pm$ (red), $B^\pm$ (blue) with conservative FST and other EIC detector efficiencies compared to models of parton energy  loss (solid lines) and in-medium hadronization (dashed lines). Projected measured shape is set to be flat for simplicity.}
\label{fig:money}
\end{figure}

Let us now turn to one physics example at the EIC. The tremendous discriminating power of heavy flavor  with respect to models of energy loss (solid lines) and hadronization in nuclear matter (dashed lines) is shown in Fig.~\ref{fig:money}. We use the soft gluon emission energy loss limit of the full splitting kernels described above. The observable 
\begin{equation}
R_{eA} = \frac{d\sigma^{eA}}{d\nu dz_h} \bigg/ A \cdot \frac{d\sigma^{ep}}{d\nu dz_h} 
\end{equation}
 is the hadron cross sections ration in $e+A$ to $e+p$ reactions, appropriately normalized, versus the hadron momentum fraction $z_h$. 
 The calculation is performed for $\nu =20$~GeV and $Q^2$ integrated in the relevant kinematic range.
  $R_{eA}$ shows strong enhancement for $D$-mesons and $B$-mesons if hadronization takes place outside of the large nucleus, where
  the heavy quark-to-meson fragmentation functions are taken from~\cite{Sharma:2009hn}.
  $R_{eA}$ is, however,   suppressed if hadrons are formed early and absorbed in nuclear matter. Such differentiation is  not possible with light hadrons (shown by solid and dashed black lines) since both models predict similar cross section attenuation.  We note that the absorption model shown here relies on the fact that  formation time of heavy hadrons decreases with mass.
  
To perform this measurement, a forward silicon tracker (FST) in the proton or nucleus going direction at the EIC is necessary. Conceptual design for such
tracker has been developed by Los Alamos National Laboratory (LANL) scientists.  Our simulations show that  with  one year of low luminosity running at center-of-mass energy  
$\sqrt{s}=69.3$~GeV, the statistical precision of the measurement is  high, as illustrated by the small statistical error bars in Fig.~\ref{fig:money}. Future heavy flavor results from the EIC will allow us to  pin down the opacities of cold nuclear matter and transport properties of large nuclei.  For reference, the yellow band in  Fig.~\ref{fig:money} shows the effect of a factor of two uncertainty in nuclear opacity  on the  distribution of $D$-mesons in $e+A$ collisions. It is clear that data taken with the FST can constrain such quantities at the $ \sim20\%$ level.  Decorrelation of hadrons due to multiple scattering in nuclear matter~\cite{Qiu:2003pm} can complement cross section attenuation measurements.

\section{Conclusions}
In summary, inclusive hadron and jet production  will play an important role at the EIC in determining the transport properties of nuclei. With this in mind,  we presented solutions for the $q \rightarrow qg$, $g\rightarrow gg$, $q \rightarrow g q$, $g \rightarrow q\bar{q}$  in-medium splitting kernels  beyond the soft gluon approximation  to  ${\cal O}(\alpha_s)$ and to  any desired order in opacity.
We also demonstrated that the full process-dependent in-medium splitting kernels can be evaluated numerically to higher orders in opacity in a  realistic nuclear  medium. The splitting functions are ready for broad phenomenological applications, such as jet and heavy flavor production at the EIC.  
 
 The way in which the cross sections of heavy mesons are modified in reactions with  nuclei of known size (such as  Cu or  Au) relative to the ones measured in reactions with protons provides an experimental handle on the space-time dynamics of hadronization -- the process that describes how elementary particles are formed from the fundamental quark and gluon constituents.  The way in which heavy quarks lose energy as they propagate in the nucleus is a key diagnostic of the dynamic response and not yet known transport properties of nuclear matter. Knowledge of such properties is essential to understand particle transport in extremely dense environments, for example neutron stars. For these measurements to become reality, 
 silicon tracking with high spatial and timing resolution at the EIC is necessary.


\begin{thebibliography}{1}

\bibitem{Airapetian:2007vu}
A.~Airapetian {\em et~al.}, {Hadronization in semi-inclusive deep-inelastic
  scattering on nuclei}, {\em Nucl. Phys.} {\bf B780}, 1  (2007).

\bibitem{Kopeliovich:2003py}
B.~Z. Kopeliovich, J.~Nemchik, E.~Predazzi and A.~Hayashigaki, {Nuclear
  hadronization: Within or without?}, {\em Nucl. Phys.} {\bf A740}, 211
  (2004).

\bibitem{Arleo:2003jz}
F.~Arleo, {Quenching of hadron spectra in DIS on nuclear targets}, {\em Eur.
  Phys. J.} {\bf C30}, 213  (2003).

\bibitem{Kang:2016ofv}
Z.-B. Kang, F.~Ringer and I.~Vitev, {Effective field theory approach to open
  heavy flavor production in heavy-ion collisions}, {\em JHEP} {\bf 03}, p. 146
   (2017).

\bibitem{Sievert:2018imd}
M.~D. Sievert and I.~Vitev, {Quark branching in QCD matter to any order in
  opacity beyond the soft gluon emission limit}, {\em Phys. Rev.} {\bf D98}, p.
  094010  (2018).

\bibitem{Sievert:2019cwq}
M.~D. Sievert, I.~Vitev and B.~Yoon, {A complete set of in-medium splitting
  functions to any order in opacity}, {\em Phys. Lett.} {\bf B795}, 502
  (2019).

\bibitem{Sharma:2009hn}
R.~Sharma, I.~Vitev and B.-W. Zhang, {Light-cone wave function approach to open
  heavy flavor dynamics in QCD matter}, {\em Phys. Rev.} {\bf C80}, p. 054902
  (2009).

\bibitem{Qiu:2003pm}
J.-w. Qiu and I.~Vitev, {Transverse momentum diffusion and broadening of the
  back-to-back dihadron correlation function}, {\em Phys. Lett.} {\bf B570},
  161  (2003).

\end{thebibliography}

\end{document}